\documentclass[11pt]{article}
\usepackage{graphicx}
\begin{document}

\title{Condensate Fraction and Pair Coherence Lengths of Two-Dimension Fermi Gases with Spin-Orbit Coupling}

\author{BeiBing Huang\thanks{Corresponding author.
Electronic address: hbb4236@mail.ustc.edu.cn}\\
Department of Experiment Teaching, Yancheng Institute of Technology,
Yancheng, 224051, China
\\ and Shaolong Wan\\
Institute for Theoretical Physics and Department of Modern Physics \\
University of Science and Technology of China, Hefei, 230026, China}

\maketitle
\begin{center}
\begin{minipage}{120mm}
\vskip 0.8in
\begin{center}{\bf Abstract} \end{center}

{The effects of Rashba spin-orbit coupling on BCS-BEC crossover, the
condensate fraction and pair coherence lengths for a two-component
attractive Fermi gas in two dimension are studied. The results at
$T=0K$ indicate that (1) when the strength of SOC is beyond a
critical value, BCS-BEC crossover does not happen in a conventional
sense; (2) SOC enhances the condensate fraction, but suppresses pair
coherence lengths.}

\end{minipage}
\end{center}

\vskip 1cm

\textbf{PACS} number(s):03.75.Ss, 05.30.Fk, 67.85.Lm
\\

In a crystalline solid spin-orbit coupling (SOC), which occurs
naturally in systems with broken inversion symmetry and makes the
spin degree of freedom respond to its orbital motion, is responsible
for many interesting phenomena, such as magnetoelectric effect
\cite{ide, kato, ganichev}, visionary Datta-Das spin transistor
\cite{datta, jsch}, topological insulator \cite{hasan, kane} and
superconductivity \cite{ludwig, sarma}. Taking topological
superconductivity for example, it has been predicted to occur in
superconductors with a sizable spin-orbit coupling in the presence
of an external magnetic field \cite{sau1, sau2, sau3, sau4, sato}.
In these systems the transition to topological phases requires that
critical magnetic field is much larger than the superconductivity
gap above which an s-wave superconductor is expected to vanish in
the absence of SOC. It is SOC that competes with a strong magnetic
field to give rise to a topological superconducting phase.

It is widely known that ultracold atom systems can be used to
simulate many other systems owing to their many controllable
advantages and operabilities \cite{greiner, moritz, lewenstein}.
Certainly the simulations to SOC, which are generally equivalent to
produce non-abelian gauge potential with optical \cite{op1, op2,
op3} or radio-frequency fields \cite{rf}, are also possible and have
been realized in a neutral atomic Bose-Einstein condensate (BEC) by
dressing two atomic spin states with a pair of lasers
\cite{spileman}. Motivated by such a pioneer experiment and a
practical proposal for generating SOC in ${}^{40}K$ atoms
\cite{k40}, BCS-BEC crossover in the two-component Fermi gases with
SOC have been widely studied \cite{cross3, cross2, cross1, vbshenoy,
melo2, chengang, cross4, melo}. On the one hand for balanced case
SOC not only leads to an anisotropic superfluid \cite{cross3}, the
signatures of which could be observed in the momentum distribution
or the single-particle spectral function of atomic cloud, but also
significantly enhances the superfluid transition temperature when
scattering length $a_s<0$, while suppresses it slightly when $a_s>0$
\cite{cross2}. In addition by adjusting the strength of SOC, one can
engineer a BCS-BEC crossover even with a very weak attractive
interaction that is unable to produce a two-body bound-state in free
vacuum \cite{cross1}. On the other hand for imbalanced case SOC and
population imbalance are counteracting, and this competition tends
to stabilize the uniform superfluid phase against the phase
separation. However, SOC stabilizes (destabilizes) the uniform
superfluid phase against the normal phase for low (high) population
imbalances \cite{cross4}.

In this paper we consider BCS-BEC crossover of two-component Fermi
gases with SOC positioned in a two-dimensional (2D) space, and are
interested in the evolutions of condensate fraction and pair
coherence lengths along the crossover. As is known to all that,
without SOC, a spin-up fermion pairs with a spin-down fermion, i.e.
fermion pairs happen in the singlet channel. This leads to that the
condensate fraction comes from the contribution of singlet pairs
\cite{momem, luca} and the coherence length of singlet pairs is
defined \cite{prb2, prb1}. In the presence of SOC, SOC also induces
triplet pairs in the system besides singlet pairs. Thus at this time
both singlet and triplet pairs contribute to the condensate
fraction, and we must define two coherence lengths related to
singlet and triplet pairs.

The Hamiltonian of the system we consider is
\begin{eqnarray}
H=\int d^2\vec{r}\left\{\sum_{\alpha, \gamma}
\Psi^{\dag}_{\alpha}(\vec{r})\left[\frac{\vec{p}^2}{2m} +  v_R
(\sigma_x p_y - \sigma_y p_x)
-\mu\right]\Psi_{\gamma}(\vec{r})-U\Psi^{\dag}_{\uparrow}(\vec{r})
\Psi^{\dag}_{\downarrow}(\vec{r}) \Psi_{\downarrow}(\vec{r})
\Psi{\uparrow}(\vec{r})\right\},  \label{1}
\end{eqnarray}
where a Fermi atom of mass $m$ for spin $\alpha$ is described by the
field operator $\Psi_{\alpha}(\vec{r})$. $\sigma_x$ and $\sigma_y$
denote the Pauli matrices in the $x$ and $y$ directions. $-U(U>0)$
corresponds to attractive contact interaction among fermions and
$\mu$ is the chemical potential. Without loss of generality we
assume SOC to be Rashba type and its strength to be $v_R$.
Transforming the field operator $\Psi_{\alpha}(\vec{r})$ into
momentum space
\begin{eqnarray}
\Psi_{\alpha}(\vec{r})=\frac{1}{\sqrt{V}}\sum_k
\Psi_{\alpha}(\vec{k})e^{i \vec{k}\cdot\vec{r}}, \label{2}
\end{eqnarray}
where $V$ is the volume of the system, and introducing the
mean-field order parameter $\Delta =
\frac{U}{V}\sum_k<\Psi_{\downarrow}(-\vec{k})
\Psi_{\uparrow}(\vec{k})>$ the Hamiltonian (\ref{1}) is written into
\begin{eqnarray}
H&=&\sum_{k} \xi_k[\Psi_{\uparrow}^{\dag}(\vec{k})
\Psi_{\uparrow}(\vec{k})+\Psi_{\downarrow}^{\dag}(\vec{k})
\Psi_{\downarrow}(\vec{k})]+\varrho_k\Psi_{\uparrow}^{\dag}(\vec{k})
\Psi_{\downarrow}(\vec{k})+\varrho_k^*\Psi_{\downarrow}^{\dag}(\vec{k})
\Psi_{\uparrow}(\vec{k})\nonumber \\
&&-\Delta\Psi_{\uparrow}^{\dag}(\vec{k})\Psi_{\downarrow}^{\dag}(-\vec{k})
-\Delta\Psi_{\downarrow}(-\vec{k})\Psi_{\uparrow}(\vec{k})
+\frac{V}{U}\Delta^2 \label{3}
\end{eqnarray}
with $\xi_k=\frac{\hbar^2k^2}{2m}-\mu$ and $\varrho_k=\hbar
v_R(k_y+ik_x)$.

Explicitly the Hamiltonian (\ref{3}) is second order about field
operators and can be solved exactly. To attack this goal we choose
to use imaginary time Green function method \cite{mahan} since in
this frame some interesting physical quantities, such as atom number
and order parameter, can be directly deduced from Green functions.
Defining two normal Green functions
$G_{\uparrow\uparrow}(\tau)=-<T_{\tau}\Psi_{\uparrow}(k, \tau)
\Psi_{\uparrow}^{\dag}(k, 0)>$,
$G_{\downarrow\uparrow}(\tau)=-<T_{\tau}\Psi_{\downarrow}(k, \tau)
\Psi_{\uparrow}^{\dag}(k, 0)>$ and two anomalous Green functions $
F_{\downarrow\uparrow}(\tau)=-<T_{\tau}\Psi_{\downarrow}^{\dag}(-k,
\tau) \Psi_{\uparrow}^{\dag}(k, 0)>$, $
F_{\uparrow\uparrow}(\tau)=-<T_{\tau}\Psi_{\uparrow}^{\dag}(-k,
\tau) \Psi_{\uparrow}^{\dag}(k, 0)>$, and using the time evolution
of an imaginary time operator $O$, $-\hbar\frac{\partial}{\partial
\tau} O =[O, H]$, we obtain the equation of motion
\begin{eqnarray}
M_{4\times4}(\overline{G}_{\uparrow\uparrow},
\overline{G}_{\downarrow\uparrow},
\overline{F}_{\downarrow\uparrow},
\overline{F}_{\uparrow\uparrow})^T=(1, 0, 0, 0)^T \label{4}
\end{eqnarray}
where $\overline{G}_{\uparrow\uparrow},
\overline{G}_{\downarrow\uparrow},
\overline{F}_{\downarrow\uparrow}, \overline{F}_{\uparrow\uparrow}$
are Fourier transformation of $G_{\uparrow\uparrow}(\tau),
G_{\downarrow\uparrow}(\tau), F_{\downarrow\uparrow}(\tau),
F_{\uparrow\uparrow}(\tau)$ respectively and
\begin{eqnarray}
M_{4\times4}&=&\left(
\begin{array}{cccc}
iw_n-\xi_k & -\varrho_k  & \Delta  &0\\
\smallskip
 -\varrho_k^* & iw_n-\xi_k & 0 & -\Delta\\
\smallskip
\Delta&0&iw_n+\xi_k&-\varrho_k\\
\smallskip
 0    &  -\Delta    &  -\varrho_k^*  & iw_n+\xi_k
\end{array}
\right), \label{5}
\end{eqnarray}
with $w_n=\frac{(2n+1)\pi}{\beta \hbar}$ representing the Matsubara
frequency. From (\ref{5}) the quasiparticle excitation spectrum is
obtained by replacing $iw_n$ with $w$ and setting det$M_{4\times4} =
0$, leading to $E_{k,\mp}^2=(\xi_k\mp|\varrho_k|)^2+\Delta^2$.
Naturally for $v_R=0$, the excitation spectrum reduces to the
standard BCS form $E_k^2=\xi_k^2+\Delta^2$. By using $E_{k,\mp}$,
Green functions can also be figured out
\begin{eqnarray}
\overline{G}_{\uparrow\uparrow}=\frac{(iw_n+\xi_k)[(iw_n)^2-\xi_k^2-\Delta^2]-(iw_n-\xi_k)|\varrho_k|^2}
{(iw_n-E_{k,-})(iw_n+E_{k,-})(iw_n-E_{k,+})(iw_n+E_{k,+})},\nonumber\\
\overline{F}_{\downarrow\uparrow}=\frac{\Delta[\xi_k^2+\Delta^2+|\varrho_k|^2-(iw_n)^2]}
{(iw_n-E_{k,-})(iw_n+E_{k,-})(iw_n-E_{k,+})(iw_n+E_{k,+})},\nonumber\\
\overline{F}_{\uparrow\uparrow}=\frac{-2\Delta\xi_k\varrho_k^*}
{(iw_n-E_{k,-})(iw_n+E_{k,-})(iw_n-E_{k,+})(iw_n+E_{k,+})}.\label{6}
\end{eqnarray}
From (\ref{6}) we know for a general $k$, Green function
$F_{\uparrow\uparrow}$ is non-zero. This signifies that SOC induces
triplet fermion pairs although the interaction is pure s-wave.

The evolution of the system is completely decided by the equations
of order parameter $\Delta$ and particle number $N$
\begin{eqnarray}
&&\frac{1}{U}=\frac{1}{V}\sum_k\left[\frac{\tanh{\frac{\beta}{2}E_{k,-}}}{4E_{k,-}}
+\frac{\tanh{\frac{\beta}{2}E_{k,+}}}{4E_{k,+}}\right] \label{7}\\
&&N=\sum_k\left[
1-\frac{\xi_k-|\varrho_k|}{2E_{k,-}}\tanh{\frac{\beta}{2}E_{k,-}}
-\frac{\xi_k+|\varrho_k|}{2E_{k,+}}\tanh{\frac{\beta}{2}E_{k,+}}\right]
\label{8}
\end{eqnarray}
by using the relations $\Delta=\frac{g}{\beta \hbar
V}\sum_{k,iw_n}\overline{F}_{\downarrow\uparrow}$ and
$N=\frac{2}{\beta
\hbar}\sum_{k,iw_n}\overline{G}_{\uparrow\uparrow}$. It is easily
found that the equation (\ref{7}) is divergent due to the fact that
we use contact interaction to simulate the true two-body potential,
whose Fourier transformation should fall off at large momentum. To
regulate the divergence the strength of contact interaction $U$
should be substituted into
\begin{eqnarray}
\frac{1}{U}=\frac{1}{V}\sum_{k}\frac{1}{\hbar^2k^2/m+\epsilon_B},
\label{9}
\end{eqnarray}
where $\epsilon_B$ is the 2D two-body binding energy
\cite{randeria}.

We self-consistently solved the equations (\ref{7}) and (\ref{8})
for the different strengths $v_R$ of SOC at $T=0K$. In Fig.1(a), the
evolution of the order parameter $\Delta$ is shown and we can find
that the larger $v_R$ is, the larger $\Delta$ is. From this
perspective the existence of SOC enhances the superfluidity of the
system. By comparison the evolution of $\mu$ is more interesting and
plotted in Fig.1(b). It is easily found that with the increase of
$v_R$, the chemical potential $\mu$ becomes negative even if the
two-body binding energy $\epsilon_B$ is very small. Hence from a
conventional viewpoint, that BCS-BEC crossover is exactly realized
when the chemical potential $\mu$ crosses zero, this fact suggests
that when $v_R$ is beyond a critical value $v_R^c$, BCS-BEC
crossover does not happen at all and the system is into BEC. From
numerical work $v_R^c\approx v_F/\sqrt{2}$, where $v_F$ is 2D Fermi
velocity and has a relation with particle density $n$
$v_F=\hbar\sqrt{2\pi n}/m$. It is to be noted that 2D Fermi energy
$\epsilon_F=n\pi\hbar^2/m$ and Fermi wavevector $k_F=\sqrt{2\pi n}$.

Now we analyze the condensate fraction of fermions in the BCS-BEC
crossover with SOC. In terms of a superfluid Fermi system the
condensate fraction $n_0$ is related to off-diagonal long-range
order (ODLRO) \cite{yang} and corresponds to the maximal eigenvalue
$N_0$, divided by the whole particle number $N$, of two-particle
density matrix
\begin{eqnarray}
\rho_2(\vec{r}_1\sigma_1,\vec{r}_2\sigma_2:\vec{r}_1^,\sigma_1^,,\vec{r}_2^,\sigma_2^,)
=<\Psi_{\sigma_1}^{\dag}(\vec{r}_1)\Psi_{\sigma_2}^{\dag}(\vec{r}_2)
\Psi_{\sigma_2^,}(\vec{r}_2^,)\Psi_{\sigma_1^,}(\vec{r}_1^,)>.
\label{10}
\end{eqnarray}
According to Leggett's book \cite{leggett}, $N_0$ can be decided as
follows
\begin{eqnarray}
N_0=\sum_{\sigma_1,\sigma_2}\int d\vec{r}_1\int d\vec{r}_2
|\Phi(\vec{r}_1\sigma_1,\vec{r}_2\sigma_2)|^2, \label{11}
\end{eqnarray}
where $\Phi(\vec{r}_1\sigma_1,\vec{r}_2\sigma_2)=
<\Psi_{\sigma_1}(\vec{r}_1)\Psi_{\sigma_2}(\vec{r}_2)>$ is an
anomalous average which arises as a result of spontaneous breaking
of the $U(1)$ gauge symmetry.

According to the above theory and a fact that SOC induces triplet
pairs, the condensate fraction in the presence of SOC is
\begin{eqnarray}
n_0&=&\frac{2}{N}\sum_k\left[|<\Psi_{\uparrow}(\vec{k})\Psi_{\downarrow}(-\vec{k})>|^2+
|<\Psi_{\uparrow}(\vec{k})\Psi_{\uparrow}(-\vec{k})>|^2\right].\label{12}
\end{eqnarray}
The pre-factor $2$ comes from the contributions of spin summation
and time-reversal symmetry. In contrast to the case without SOC,
there is an extra contribution to the condensate fraction from
triplet pairs. Following the same procedure deriving the equations
(\ref{7}) and (\ref{8}), we have
\begin{eqnarray}
n_0 &=&\frac{\Delta^2}{4N}\sum_k\left[
\frac{\tanh^2{\frac{\beta}{2}E_{k,-}}}{E_{k,-}^2}+
\frac{\tanh^2{\frac{\beta}{2}E_{k,+}}}{E_{k,+}^2}\right].\label{12'}
\end{eqnarray}
When $v_R=0$, $E_{k,-}=E_{k,+}=E_k$ and
$n_0=\frac{\Delta^2}{2N}\sum_k
\tanh^2{\frac{\beta}{2}E_{k}}/E_{k}^2$, same as the results in
\cite{luca}. In Fig.2 we calculate the condensate fraction $n_0$ for
different $v_R$ at $T=0K$. The results are twofold. Firstly SOC also
enhances the condensate fraction, which is consistent with Fig.1(a).
In addition, with the increase of $\epsilon_B$ $n_0$ rapidly
increases to a large value for a large $v_R$. Maybe this phenomenon
can be illustrated from the results of Fig.1(b), that for a large
$v_R$ there is not BCS-BEC crossover and the system is situated in
BEC, building on the fact that for a weakly interacting BEC, almost
all atoms are into the condensate. A recent paper
\cite{lucasalasnich} also calculate the condensate fraction in the
same system but do not think over the contribution from triplet
pairs, so their result is qualitatively incorrect.

At last we determine the coherence lengths for singlet and triplet
pairs. On general ground, information on pair coherence lengths can
be extracted from the pair-distribution function
\begin{eqnarray}
g_{\sigma_1,\sigma_2}(r)=\frac{1}{n^2}<\Psi_{\sigma_1}^{\dag}(\vec{r})
\Psi_{\sigma_2}^{\dag}(0) \Psi_{\sigma_2}(0)
\Psi_{\sigma_1}(\vec{r})>. \label{13}
\end{eqnarray}
Following the same spirit that the Hartree-Fock term has been
neglected in the Hamiltonian (\ref{3}), at the mean-field level
(\ref{13}) becomes
\begin{eqnarray}
g_{\sigma_1,\sigma_2}(r)=\frac{1}{n^2}|<\Psi_{\sigma_1}^{\dag}(\vec{r})
\Psi_{\sigma_2}^{\dag}(0)>|^2, \label{14}
\end{eqnarray}
and pair coherence lengths $\xi_{pair}^s$, $\xi_{pair}^t$ for
singlet and triplet pairs can be obtained as
\begin{eqnarray}
(\xi_{pair}^s)^2=\frac{\int d\vec{r}
\vec{r}^2g_{\uparrow,\downarrow}(r)}{\int
d\vec{r} g_{\uparrow,\downarrow}(r)}=\frac{\sum_k \nabla_k\varphi_k^{s*}\cdot\nabla_k\varphi_k^{s}}{\sum_k \varphi_k^{s*}\varphi_k^{s}}\label{15}\\
(\xi_{pair}^t)^2=\frac{\int d\vec{r}
\vec{r}^2g_{\uparrow,\uparrow}(r)}{\int d\vec{r}
g_{\uparrow,\uparrow}(r)}=\frac{\sum_k
\nabla_k\varphi_k^{t*}\cdot\nabla_k\varphi_k^{t}}{\sum_k
\varphi_k^{t*}\varphi_k^{t}}\label{16}
\end{eqnarray}
with
$\varphi_k^{s}=\frac{\Delta}{4}\left[\frac{\tanh{\frac{\beta}{2}E_{k,-}}}{E_{k,-}}+
\frac{\tanh{\frac{\beta}{2}E_{k,+}}}{E_{k,+}}\right]$ and
$\varphi_k^{t}=\frac{\Delta\varrho_{k}^*}{4\varrho_{k}}
\left[\frac{\tanh{\frac{\beta}{2}E_{k,-}}}{E_{k,-}}
-\frac{\tanh{\frac{\beta}{2}E_{k,+}}}{E_{k,+}}\right]$. Without SOC,
$\varphi_k^t=0$ but
$\varphi_k^s=\frac{\Delta}{2}\tanh{\frac{\beta}{2}E_{k}}/E_{k}$,
consistent with the results in \cite{prb2, prb1}. Fig.3 and Fig.4
describe the behaviors of $\xi_{pair}^s$ and $\xi_{pair}^t$ in the
process of evolution at $T=0K$, respectively. Very explicitly, SOC
suppresses pair coherence lengths for both singlet and triplet
pairs, and for a large $v_R$, pair coherence lengths rapidly
decrease to a small value. We think that this phenomenon can also be
understood from Fig.1(b) in the light of the fact in BEC region pair
coherence length is much smaller than in BCS region. Besides by
comparing such two figures, triplet pair coherence length
$\xi_{pair}^t$ always is larger than singlet pair coherence length
$\xi_{pair}^s$. Physically this is the result from Pauli exclusion
principle.

In summary we have discussed the evolution from BCS to BEC
superfluids in the presence of Rashba SOC in two dimension and shown
that when the strength of SOC is beyond a critical value, BCS-BEC
crossover does not happen in a conventional sense. In addition, we
studied the evolutions of the condensate fraction and pair coherence
lengths. The results indicate that SOC enhances the condensate
fraction, but suppresses pair coherence lengths. Furthermore we also
give some physical interpretation for some phenomena.

\section*{Acknowledgement}

The work was supported by National Natural Science Foundation of
China under Grant No. 10675108 and Foundation of Yancheng Institute
of Technology under Grant No. XKR2010007.

\begin{figure}[htbp]
\includegraphics{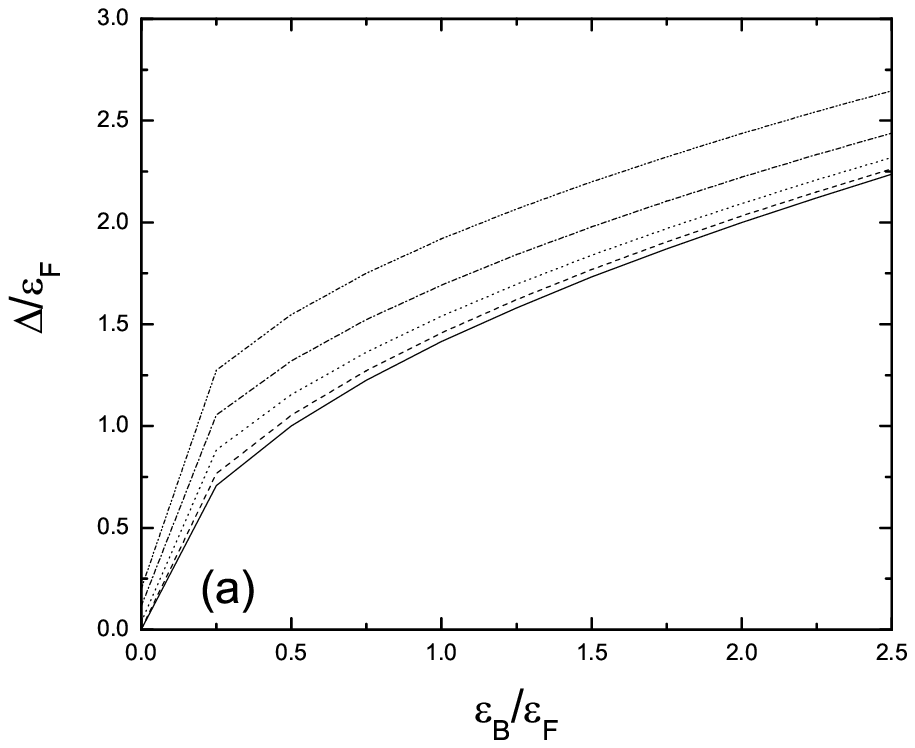}
\includegraphics{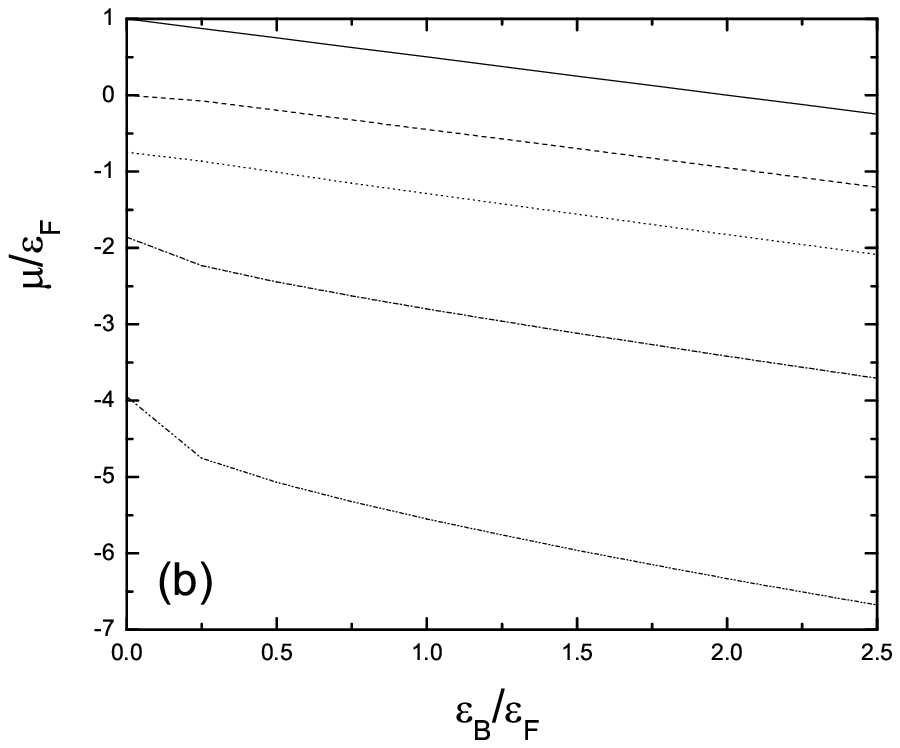}
\caption{The evolutions of order parameter $\Delta$ (a) and chemical
potential $\mu$ (b) as a function of two-body binding energy
$\epsilon_B$, in units of the Fermi energy $\epsilon_F$, for
different the strengths $v_R$ of SOC at $T=0K$. In (a) from bottom
to top and in (b) from top to bottom $v_R/v_F=0, \sqrt{2}/2, 1,
\sqrt{2}, 2$.} \label{fig.1}
\end{figure}

\begin{figure}[htbp]
\includegraphics{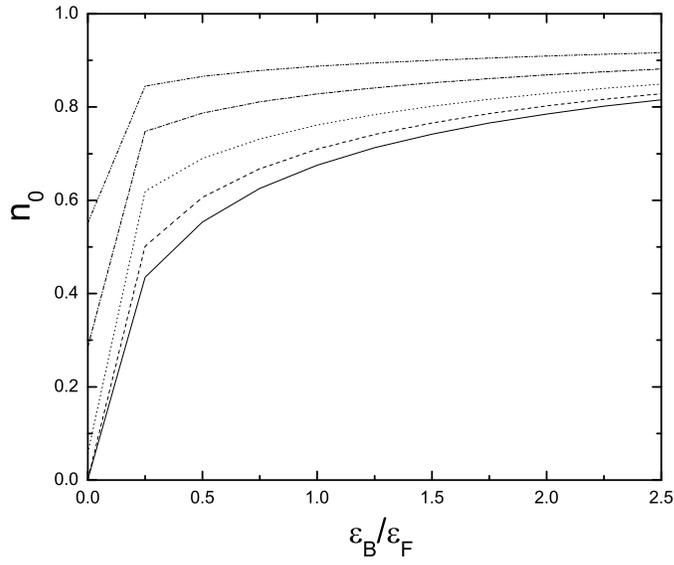}
\caption{The evolution of the condensate fraction $n_0$ as a
function of two-body binding energy $\epsilon_B$, in units of the
Fermi energy $\epsilon_F$, for different the strengths $v_R$ of SOC
at $T=0K$. From bottom to top $v_R/v_F=0, \sqrt{2}/2, 1, \sqrt{2},
2$.} \label{fig.2}
\end{figure}

\begin{figure}[htbp]
\includegraphics{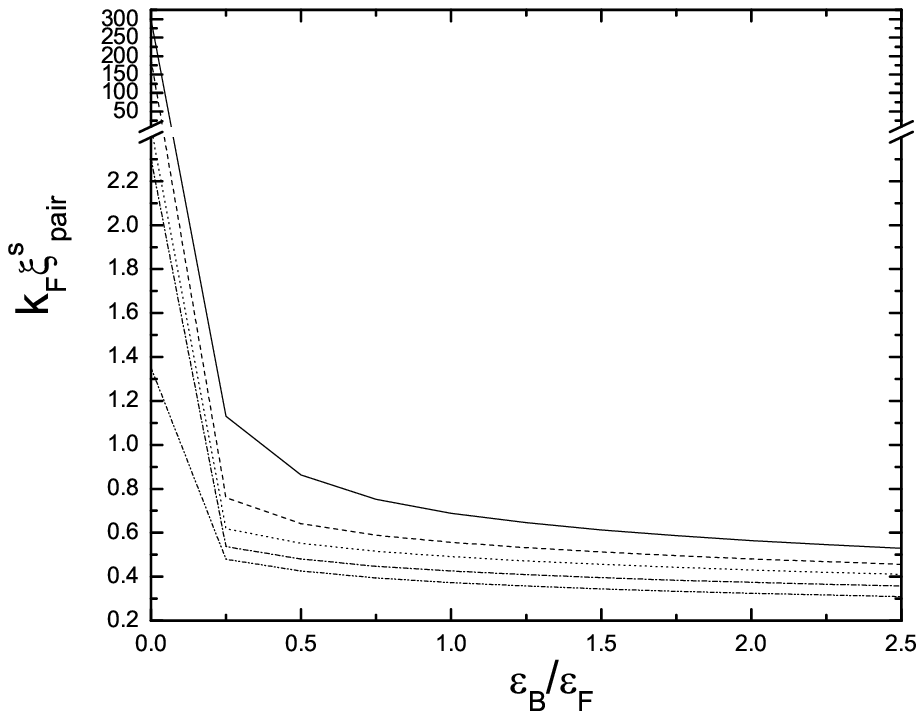}
\caption{The evolution of singlet pairs coherence length
$\xi_{pair}^s$ as a function of two-body binding energy
$\epsilon_B$, in units of the inverse Fermi wavevector $k_F$ and
Fermi energy $\epsilon_F$ respectively, for different the strengths
$v_R$ of SOC at $T=0K$. From top to bottom $v_R/v_F=0, \sqrt{2}/2,
1, \sqrt{2}, 2$.} \label{fig.3}
\end{figure}

\begin{figure}[htbp]
\includegraphics{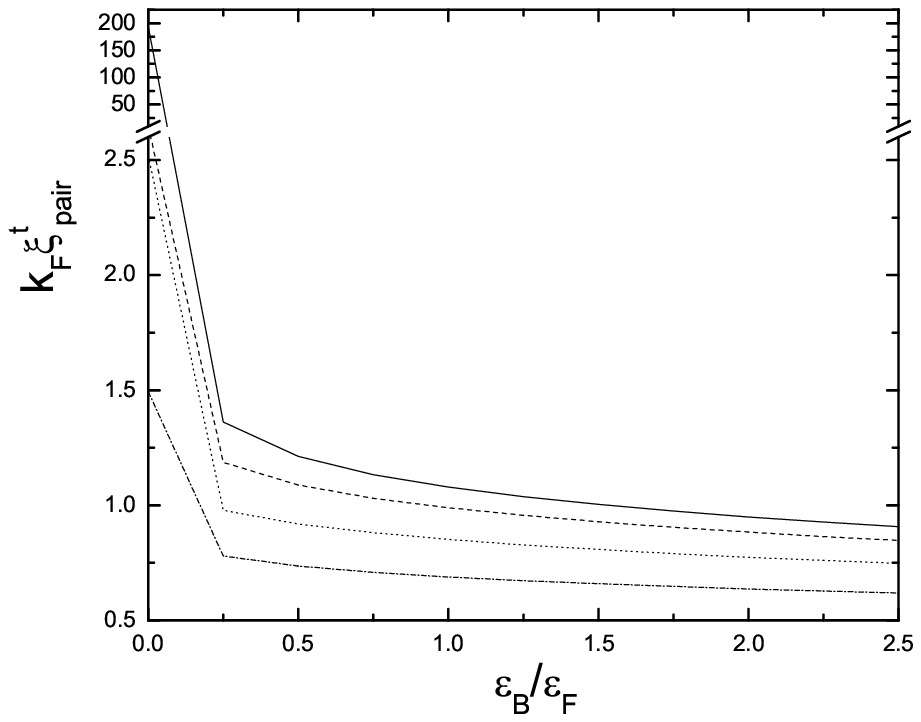}
\caption{The evolution of triplet pairs coherence length
$\xi_{pair}^t$ as a function of two-body binding energy
$\epsilon_B$, in units of the inverse Fermi wavevector $k_F$ and
Fermi energy $\epsilon_F$ respectively, for different the strengths
$v_R$ of SOC at $T=0K$. From top to bottom $v_R/v_F=\sqrt{2}/2, 1,
\sqrt{2}, 2$.} \label{fig.4}
\end{figure}

\end{document}